\documentclass[prd,aps,showpacs,preprintnumbers,amssymb]{revtex4}
\usepackage{graphicx}

\def\e3p{$\eta \rightarrow 3 \pi$}

\begin{document}

\title{%
\hfill{\normalsize\vbox{%
\hbox{\rm SU-4252-915}
 }}\\
{Are three flavors special?}}

\author{Amir H. Fariborz $^{\it \bf a}$~\footnote[1]{Email:
 fariboa@sunyit.edu}}

\author{Renata Jora $^{\it \bf b}$~\footnote[2]{Email:
 rjora@ifae.es}}

\author{Joseph Schechter $^{\it \bf c}$~\footnote[3]{Email:
 schechte@physics.syr.edu}}

\author{M. Naeem Shahid $^{\it \bf c}$~\footnote[4]{Email:
 mnshahid@physics.syr.edu}}

\affiliation{$^ {\bf \it a}$ Department of Engineering, Science
 and Mathematics,
 State University of New York Institute of Technology, Utica,
 NY 13504-3050, USA.}

\affiliation{$^ {\bf \it b}$ Grup de Fisica Teorica and IFAE,
Universitat Autonoma de Barcelona, E-08193 Bellaterra (Barcelona),
Spain.}

\affiliation{$^ {\bf \it c}$ Department of Physics,
 Syracuse University, Syracuse, NY 13244-1130, USA,}

\date{\today}

\begin{abstract}

    It has become clearer recently that the regular pattern
of three flavor nonets describing the low spin meson multiplets
seems to require
some modification for the case of the spin 0 scalar mesons. One picture
which has
 had some success, treats the scalars in a chiral Lagrangian
framework and considers them to populate two nonets. These are,
in turn,  taken to
result from the mixing of two ``bare" nonets, one of which is of quark-
antiquark type and the other of two quark- two antiquark type. Here we
show that such a mixing is, before chiral symmetry breaking terms are
included, only possible for three flavors. In other cases, the two types
of structure can not have the same chiral symmetry transformation
property. Specifically, our criterion would lead one to believe that
scalar and pseudoscalar states containing charm would not have "four
quark" admixtures.
 This work is of potential interest for constructing chiral
Lagrangians based on exact chiral symmetry which is then broken by
well known specific terms. It may also be of interest in
studying some kinds of technicolor theories.

\end{abstract}


\maketitle

\section{Introduction}

    Historically, the nonet structure of elementary particle
multiplets has suggested the spin 1/2 quark substructure and,
with the help of
the "slightly" broken flavor symmetry SU(3), has provided an enormous
amount
of information
about the properties of the observed low lying hadronic states. For
example, the
lightest meson multiplets appear to be those of the pseudoscalars and
vectors,
consistent with s-wave quark- antiquark bound states. The next heaviest
set of meson multiplets seems to be generally consistent with p-wave
bound
states, yielding a scalar nonet, a tensor nonet and two axial vector
nonets.

    Available evidence indicates that the predicted states arising from
the addition of the charm and beauty quarks would fit in with
corresponding SU(4) and SU(5) extensions (having respectively
16 and 25 members) of the SU(3) nonets. Of course a possible extension
to states made with top quarks is of less interest, owing to the rapid
weak decay of the top quark.
Naturally the much heavier masses of the c and b quarks make the SU(4)
 and SU(5) symmetries not as good as SU(3). Nevertheless the observed
particles
still fit into the extended multiplets.

   However, in the last few years there has been a growing recognition
\cite{pdg} - \cite{CCL} that the
lightest nine scalar states do not seem to fit well into the above
classification. In terms of increasing mass these comprise the isosinglet
$\sigma$(600), the two isodoublets $\kappa$(800) and the roughly
degenerate
isosinglet $f_0$(980) and isotriplet $a_0$(980). There are two unexpected
features. First the masses of these states are
significantly lower than the other "constituent quark model ``p-wave
states" (i.e. tensors and two axial vectors with different C properties).
 Secondly, the order, with increasing mass - isosinglet, isodoublet
and roughly degenerate isosinglet with isotriplet - seems to be reversed
compared
to that of the ``standard" vector meson nonet.

 Clearly such a light and reversed order nonet requires some rethinking of
the standard picture of the scalar mesons.
Actually, a long time ago,
it was observed \cite{j} that the reversed order could be explained if
the light scalar nonet were actually composed of two quarks and two
antiquarks. In that case the number of strange quarks (which determines
the direction of
increasing masses)
rises with the reversed order given. For example the lowest
mass ``isolated" isosinglet scalar $\sigma$(600) would
look like
$(u{\bar u}+d{\bar d})^2$ while, for comparison, the highest mass isolated
vector isosinglet  $\phi$(1020)
looks like $s{\bar s}$.
At that time the existence of a light sigma and a light kappa was
considered dubious. More recent work by a great many people
has
now pretty much confirmed the existence of such states as well as the
plausibility that they fit into a three flavor nonet.

    The above does not necessarily explain the unusual lightness of such
a scalar nonet. A possible explanation was advanced \cite{BFS3} which
proposed that the mixing between a $qq{\bar q}{\bar q} $ scalar nonet
together with a usual p-wave
 $q{\bar q}$ nonet could produce this effect due to the "level
repulsion" expected in quantum mechanics perturbation theory.
This was done using a non-linear realization of chiral symmetry.
Further work, to be discussed below, makes use of linear realizations.
Related models for thermodynamic
properties of QCD have been presented in ref. \cite{thermo}.

\section{Two chiral nonets}

    Of course, it has been realized for a long time that the
nonet structure of mesons with respect to SU(3) flavor transformations
should, at a more
 fundamental level, be expanded to SU(3) chiral symmetry transformations;
this
amounts to an SU(3) for massless left-handed quarks and another SU(3)
for massless right-handed quarks. This chiral symmetry is that of the
 fundamental QCD Lagrangian itself, with neglect of quark mass terms.
It is accepted, for example, that the SU(3) baryon properties do not
depend
much on the u, d and s quark masses.
 The spontaneous breaking of this
symmetry, which gives zero mass pseudoscalars, is a basic part of the
present
understanding of low energy QCD. The
light quark mass
terms play a relatively small role and are treated as perturbations.
 It thus appears that
chiral (rather than just the vector) symmetry should be considered the
first approximation for an understanding of the structure
of hadrons.

This chiral point of view may be especially
relevant for studying the light scalars since they are the ``chiral
partners" of the zero mass pseudoscalars.
 To implement this picture systematically
one may
introduce
a $q{\bar q}$ chiral nonet containing 9 scalar and 9 pseudoscalar fields
 as well as a $qq{\bar q}{\bar q}$
nonet also containing 9 scalars and 9 pseudoscalars. Furthermore, the
light quark mass terms should be added as well as suitable terms to
mock up the U(1)$_A$ anomaly of QCD.

 Even though one can not write down the exact QCD wave
 functions
of the low lying mesons it is easy to write down schematic
descriptions of how quark fields may combine to give particles
with specified transformation properties.
  The usual chiral nonet $M(x)$ realizing the $q \bar q$ structure is
 then written as:

\begin{equation}
M_a^{\dot{b}} = {\left( q_{bA} \right)}^\dagger \gamma_4 \frac{1 +
\gamma_5}{2} q_ {aA}, \label{M}
\end{equation}
where $a$ and $A$ are respectively flavor and color indices. For clarity,
on the left hand side the undotted index transforms under the left SU(3)
while the dotted index transforms under the right SU(3).

    One possibility for the  $qq{\bar q}{\bar q}$ states is to make them
 as ``molecules''
 from two quark-antiquark nonets. This leads to
 the following schematic form:
\begin{equation}
M_a^{(2) \dot{b}} = \epsilon_{acd} \epsilon^{\dot{b} \dot{e} \dot{f}}
 {\left( M^{\dagger} \right)}_{\dot{e}}^c {\left( M^{\dagger}
 \right)}_{\dot{f}}^d.
\label{M2}
\end{equation}

   Note that the fields $M$ and $M^{(2)}$ transform in the same way
under chiral SU(3) as well as under the discrete P and C symmetries,
as required if they are to mix with each other according to the scheme
shown above. As noted in the Appendix, the axial U(1) transformation
properties of $M$ and $M^{(2)}$ differ from each other and
provide a measure of whether the state is of one quark-antiquark
type, two quark - antiquark type etc. In the chiral Lagrangian there
are terms which break the axial U(1) in a manner dictated by the
QCD axial anomaly.
In the Appendix it
is also pointed out that schematic fields $M^{(3)}$ and $M^{(4)}$
which have  ``diquark-antidiquark" forms instead of the ``molecular" form
can also be constructed.
There has been some discussion in the literature
about which type is
favored \cite{kknhh}.
 In the present approach either is allowable. In fact it was
shown in the first of \cite{FJS08} that the molecular form can be transformed using Fierz
identities to
a linear combination of the ``diquark-antidiquark" forms. We thus assume
that
some unspecified linear combination of  $M^{(2)}$, $M^{(3)}$ and
$M^{(4)}$,
denoted by $M^\prime$, represents the $qq{\bar q}{\bar q}$ chiral nonet
which mixes with $M$. The decomposition into pseudoscalar and scalar
fields is given by,

\begin{equation}
M=S+i\phi, \quad  M^\prime= S^\prime +i\phi^\prime.
\label{defm}
 \end{equation}

The initial discussion of the chiral Lagrangian using these
fields was presented in \cite{sectionV}. A more detailed picture
 with a particular
choice of interaction terms was given
in the first of \cite{NR04}. In a series of papers,
the model was explored for
an arbitrary choice of interactions \cite{FJS05}, a choice
 of interactions based on including terms containing less than
 a fixed number
  of underlying quark or antiquark fields \cite{1FJS07}
   and the zero quark mass limit \cite{2FJS07}.
  In addition the modeling of the axial anomaly was discussed \cite{FJS08} as
  well as the details of pion pion scattering \cite{3FJS07}.
In the most recent of these papers \cite{FJS09}, the possible
identification with all observed states was studied in further detail;
after mixing there are two physical scalar nonets and two
 physical pseudoscalar nonets. Since each nonet has one isovector,
two conjugate isospinors, and two isosinglets, there are
altogether sixteen different masses involved. The model has eight
inputs so the other eight masses are predictions. There are in fact
experimental
states which are candidates for identification with all the particles of
the model and the agreement is reasonable. Additional
predictions are given for the 4 $\times$ 4 orthogonal matrices which mix
each of the four isosinglet scalars and each of the four isosinglet
pseudoscalars. Perhaps, most interestingly, the lighter scalar mesons
are predicted to be mainly of two quark - two antiquark type while
the heavier scalar mesons are mainly of quark - antiquark type.
The situation is opposite, as expected, for the pseudoscalar mesons,
where the lighter ones are mainly of quark - antiquark type.

\section{Other than three flavors}

    Our initial motivation for this work was the recent experimental
discovery \cite{Cleo} of the semileptonic decay mode,

\begin{equation}
D_s^{+}(1968) \rightarrow f_0(980) e^{+} \nu_e,
\label{dsdecay}
\end{equation}

in which the $f_0(980)$ was identified from its two pion decay mode.
This provides some motivation for formulating a four flavor version of the
model so that the charmed meson $D_s$ would be conveniently contained.

    There is no problem finding a chiral formulation for a
 $q{\bar q}$ 16-plet, $M_a^{\dot{b}}$. However we can not find a
suitable schematic meson wave function with the same chiral
transformation property constructed, for example, as a ``molecule" out of
two such
states. The closest we can come for a two-part ``molecule" is:

\begin{equation}
M_{ag}^{(2) \dot{b} \dot{h}} = \epsilon_{agcd}
 \epsilon^{\dot{b}\dot{h} \dot{e} \dot{f}}
 {\left( M^{\dagger} \right)}_{\dot{e}}^c {\left( M^{\dagger}
 \right)}_{\dot{f}}^d.
\label{trial}
\end{equation}

However, instead of transforming under SU(4)$_L$ $\times$
 SU(4)$_R$
as $(L,R)$ = $(4,{\bar 4})$
as desired, this object transforms as  $(L,R)$ = $(6,{\bar 6})$,
owing to the two sets of antisymmetric indices
($ag$ and $\dot{b}\dot{h}$) which appear.
Hence, it should not mix in the chiral symmetry limit with the initial
four
flavor $q {\bar q}$ state. (See Eq.(\ref{M}))
Of course it would be possible to multiply the right hand side
of Eq.(\ref{trial}) by a third field
${\left( M^{\dagger} \right)}_{\dot{h}}^g$.
 That does give the correct transformation
property to mix with the four flavor version of Eq.(\ref{M}).
 However it corresponds to
a three quark- three antiquark
molecule. We assume that, especially after quark
 mass terms are added, an ``elementary particle"
 state of such a form
is unlikely to be bound.

     The same problem emerges in the four flavor case
 when we alternatively construct composites of the diquark -
antidiquark states given in Eqs.(\ref{M3}) and (\ref{M4})
 of the Appendix. As above, this yields a composite state
transforming like $(6,{\bar 6})$
 (rather than the desired $(4,{\bar 4}))$:

\begin{equation}
M_{gp}^{(3) \dot{f} \dot{q}} = {\left( L^{gpE}\right)}^\dagger
R^{\dot{f} \dot{q} E},
\label{M3try}
\end{equation}

where,

\begin{eqnarray}
L^{gpE} = \epsilon^{gpab} \epsilon^{EAB}q_{aA}^T C^{-1} \frac{1 +
 \gamma_5}{2} q_{bB}, \nonumber \\
R^{\dot{f} \dot{q} E} = \epsilon^{\dot{f} \dot{q} \dot{c} \dot{d}}
\epsilon^{EAB}q_{\dot{c}A}^T C^{-1} \frac{1 -
\gamma_5}{2}
 q_{\dot{d}B}.
\label{LRtry}
\end{eqnarray}

We could contract $L^{gpE}$ with a left handed quark field
and $R^{\dot{f} \dot{q} E}$ with a right handed quark field to
obtain the desired overall transformation property at the expense of
having a
three quark- three antiquark state (which we are assuming to be unbound).

    It is clear that essentially the same argument would hold for
 five or more quark flavors.

 Going in the direction of fewer flavors, we now
note that there is also
no suitable  schematic "molecular" wavefunction
 available in the 2-flavor case
for mixing with the quark- antiquark state. The closest we can come here
for a "molecule" has the form:

\begin{equation}
M^{(2)} = \epsilon_{cd} \epsilon^{\dot{e} \dot{f}}
 {\left( M^{\dagger} \right)}_{\dot{e}}^c {\left( M^{\dagger}
 \right)}_{\dot{f}}^d.
\label{2fla}
\end{equation}

This is clearly unsatisfactory since it transforms like (1,1)
under  SU(2)$_L$ $\times$ SU(2)$_R$
rather than the (2,2) required for mixing according to our
assumed model. Actually one must be a little more careful
because
 it is well known that the object $M^{\dot b}_a$
is not irreducible under chiral transformations in
 the 2-flavor case. It may be interesting to show
 that the same result is
 obtained when this fact is taken into account. The irreducible
representations are formed by making use of the fact that
$\tau_2 M^* \tau_2$ transforms in the same way as $M$. Then
we may consider the irreducible linear combinations:

\begin{eqnarray}
\frac{1}{\sqrt{2}}(M+\tau_2 M^* \tau_2)\equiv \sigma I +
i{\bf \pi}\cdot{\bf \tau} \nonumber \\
\frac{1}{\sqrt{2}}(M-\tau_2 M^* \tau_2)\equiv i\eta I +
{\bf a}\cdot{\bf \tau},
\label{usualnames}
\end{eqnarray}
where the usual SU(2) chiral multiplet containing ${\bf \pi}$
and $\sigma$ is recognized as well as the parity reversed
one containing $\eta$ and the isovector scalar particle ${\bf a}$.
Since SU(2)$_L$ $\times$  SU(2)$_R$ is equivalent to the group
SO(4)we may consider the fields ${\bf \pi}$ and $\sigma$ as
making up an isotopic four vector, $p_\mu$ and the fields
${\bf a}$ and $\eta$ as comprising another four vector $q_\mu$.
A ``molecule" state which could mix
with, say $p_\mu$  would have to be another four vector made as
a product of   $p_\mu$ and $q_\mu$. The combination
 $p_\mu q_\mu$ is a singlet, the combination
$\epsilon_{\mu\nu\alpha\beta} p_{\alpha} q_{\beta}$ has six
components and the symmetric traceless combination has nine
components.
 This confirms that there is no allowed mixing with
a possible molecule at the chiral level in the two flavor case.

One might wonder why, if mixing is possible in the three flavor case,
it is not possible in the two flavor case, which is just a subset of the
former. The answer is already contained in Eq.(\ref{M2}). If we want to
find something that mixes with the quark- antiquark $\pi^+$ particle we
should look at the 12 matrix element. On the right hand side, one sees
that the ``molecule" field which mixes contains an extra $s {\bar s}$
pair, which is simply not present in the two flavor model.

    Thus we see that flavor SU(3) has some interesting special features
for schematically constructing bound states with well defined chiral
transformation properties.

    A possibility for the mixing of a quark antiquark state with a
different state not of
``molecular" (or more generally, two quark - two antiquark)
type, would be to consider a so called radial excitation.
For mixing with $M^{\dot b}_a$, such a state could be schematically
written as $f(\Box)M^{\dot b}_a$, where $f$ is a function of the
d'Alembertian. In this case, one would not expect the
 inverted multiplets which appear in the ``molecular" picture.

\section{Testable consequences}

       At the three flavor level the existence of two quark -
   two antiquark states, suggested by our kinematical criterion,
   seems to have some experimental support. This gave rise to mixtures
   with the original quark - antiquark states and a doubling of the
   scalar and pseudoscalar meson spectra, as discussed in detail in
   ref. \cite{FJS09}. As an example, there are two established
   low lying isovector scalars - a$_0$(980) and the a$_0$(1450)-
   rather than the single one predicted by the non-relativistic
   quark model.

       What does it mean to say that the extra states, and hence the mixing, is not allowed at the four flavor level? Clearly
       the scalar and pseudoscalar 16-plets can not be completely
       absent since they contain also the three flavor nonets.
       Thus we conclude that the kinematical criterion should imply that
       16 -9 = 7 members of the possible 16-plets for scalars and for pseudoscalars should not be doubled. The states which should not be doubled can be conveniently described using the notation of
       Eq.(\ref{defm}). The states which should not appear are:

       \begin{eqnarray}
       scalars:\quad {S^\prime}_a^4,\quad {S^\prime}_4^a,\quad {S^\prime}_4^4, \nonumber \\
       pseudoscalars:\quad {\phi^\prime}_a^4,\quad {\phi^\prime}_4^a,\quad {\phi^\prime}_4^4.
       \label{xm}
       \end{eqnarray}

       Here $a$ = 1,2,3 and the quark corresponence is 1=u, 2=d, 3=s,
       4=c. Furthermore subscripts denote the quark transformation property and the superscripts denote the antiquark transformation property.

       Clearly the excluded states are those having non-zero charm.
       It will be interesting to see whether this holds using the large amount of new data expected from LHC.

\section{Summary and discussion}

        A three flavor chiral model of scalar and pseudoscalar
    mesons as mixtures of "quark-antiquark" with "two quark-
    two antiquark" fields has previously been seen to be able to
    explain the unusual pattern of light scalar meson masses.
    That approach used a chiral SU(3)$_L$ $\times$ SU(3)$_R$
    linear sigma model which was supplemented by invariant terms
    which model the axial U(1) anomaly as well as the usual terms
    which model the quark masses. Before it was broken, the
    U(1)$_A$ quantum number distinguished the ``two quark-
    two antiquark" mesons from the "quark-antiquark" mesons.
    The starting point for the mixing was that a schematic
     two quark - two antiquark product state could be
     constructed with the same
SU(3)$_L$ $\times$ SU(3)$_R$ transformation property as the
original "quark-antiquark" state. Of course this is just a ``kinematic"
statement and does not presume to say that the dynamical binding has
been established or that large quark masses do not change this picture.

     In the present note we have shown that this kinematical
     feature in the chiral limit does not hold for
SU(n)$_L$ $\times$ SU(n)$_R$ when n is different from three.
In the case of n = 4, it was seen that three quark- three antiquark
states could have the same transformation property
 but we assumed that the 6-object bound state and
 other higher ones ( needed for still larger n)
  would be unlikely to be bound as an ``elementary particle".

     As for our initial motivation, mentioned in section III, to
      construct a 4 flavor model for studying semi-leptonic decays
      of  charmed mesons into scalar plus leptons, a kind of hybrid
       chiral model
      will be discussed elsewhere.

      We have also noted a possible experimental test of
      the kinematical criterion for the doubling of scalar and pseudoscalar states in the charm sector.

 \section*{Acknowledgments} \vskip -.5cm
 We would like to thank S. Stone for pointing out and discussing the
 CLEO results \cite{Cleo}. We are
happy to thank A. Abdel-Rehim, D. Black, M.
Harada, S. Moussa, S. Nasri and F. Sannino for
many helpful related discussions.
The work of
A.H.Fariborz has been partially supported by the NSF
Grant 0854863.
The work
of R.Jora has been supported by
CICYT-FEDEF-FPA 2008-01430. The work of
 J.Schechter and M.N. Shahid was supported in part by the U.
S. DOE under Contract no. DE-FG-02-85ER 40231.

\appendix
\section{Notation and further details}

Here we briefly discuss some notational and technical details.
 The $\gamma$
matrices and the charge conjugation matrix have the form:
\begin{eqnarray}
\gamma_i= \left[
\begin{array} {cc}
0&-i\sigma_i\\
i\sigma_i&0
\end{array}
\right], \hspace{1cm} \gamma_4=\left[
\begin{array}{cc}
0&1\\
1&0
\end{array}
\right],
\hspace{1cm} \gamma_5=\left[
\begin{array}{cc}
1&0\\
0&-1
\end{array}
\right],
\hspace{1cm}
 C= \left[
\begin{array} {cc}
-\sigma_2&0\\
0&\sigma_2
\end{array}
\right]. \label{candsigma}
\end{eqnarray}

Our convention
 for matrix notation is $M_a^b \rightarrow M_{ab}$.  Then $M$
 transforms under chiral $SU(3)_L \times SU(3)_R$, charge
 conjugation $C$ and
parity $P$ as
\begin{eqnarray}
&&M \rightarrow U_L M U_R^\dagger
\nonumber\\
&&C: \quad M \rightarrow  M^T, \quad \quad P: \quad M({\bf x})
 \rightarrow  M^{\dagger}(-{\bf x}).
\label{Mchiral}
\end{eqnarray}
Here $U_L$ and $U_R$ are unitary, unimodular matrices associated
with the
 transformations on the left handed
 ($q_L = \frac{1}{2}\left( 1 + \gamma_5 \right) q$) and right
 handed ($q_R = \frac{1}{2}\left( 1 - \gamma_5 \right) q$)
 quark projections.
 For the $U(1)_A$ transformation one has:
\begin{equation}
M \rightarrow e^{2i\nu} M. \label{MU1A}
\end{equation}
Next consider nonets with `` four quark",
 $qq{\bar q}{\bar q}$ structures.
An alternate possibility to the one given in Eq.(\ref{M2})of
section II
 is that such states may be bound states of a
diquark and an anti-diquark. There are two choices if the
 diquark is required to belong to a ${\bar 3}$ representation of
flavor SU(3). In the first case it belongs to a ${\bar 3}$
 of color
and is a spin singlet with the structure,
\begin{eqnarray}
L^{gE} = \epsilon^{gab} \epsilon^{EAB}q_{aA}^T C^{-1} \frac{1 +
 \gamma_5}{2} q_{bB}, \nonumber \\
R^{\dot{g} E} = \epsilon^{\dot{g} \dot{a} \dot{b}}
\epsilon^{EAB}q_{\dot{a}A}^T C^{-1} \frac{1 -
\gamma_5}{2}
 q_{\dot{b}B}.
\end{eqnarray}
 Then the matrix $M$ has the form:
\begin{equation}
M_g^{(3) \dot{f}} = {\left( L^{gA}\right)}^\dagger R^{fA}. \label{M3}
\end{equation}
  In a second alternate possibility,
 the diquark belongs to a $6$ representation of color and has spin 1.
  It has the schematic chiral realization:
\begin{eqnarray}
L_{\mu \nu,AB}^g = L_{\mu \nu,BA}^g = \epsilon^{gab}
 q^T_{aA} C^{-1}
 \sigma_{\mu \nu} \frac{1 + \gamma_5}{2} q_{bB}, \nonumber \\
R_{\mu \nu,AB}^{\dot{g}} = R_{\mu \nu,BA}^{\dot{g}} =
\epsilon^{\dot{g} \dot{a} \dot{b}}
 q^T_{\dot{a}A} C^{-1}
 \sigma_{\mu \nu} \frac{1 - \gamma_5}{2} q_{\dot{b}B},
\end{eqnarray}
where $\sigma_{\mu \nu} = \frac{1}{2i}
 \left[ \gamma_\mu, \gamma_\nu
\right]
 $.  The corresponding $M$ matrix has the form
\begin{equation}
M_g^{(4) \dot{f}} = {\left( L^{g}_{\mu \nu,AB}\right)}^\dagger
 R^{f}_{\mu
\nu,AB},
\label{M4}
\end{equation}
where the dagger operation includes a factor
 ${(-1)}^{\delta_{\mu 4}
+
 \delta_{\nu 4}}$.The nonets $M^{(2)}$, $M^{(3)}$ and
 $M^{(4)}$ transform
  like $M$
 under all of $SU(3)_L \times SU(3)_R$, $C$, $P$. Under $U(1)_A$
 all three transform with the phase $e^{-4 i \nu}$, e.g.:
\begin{equation}
M^{(2)} \rightarrow e^{-4i\nu} M^{(2)}.
\end{equation}
     It is seen that the $U(1)_A$
transformation distinguishes the ``four quark" from the
``two quark" states. In the full chiral Lagrangian treatment of the model
under
discussion there are explicit terms which model the breaking of
this symmetry and hence cause the mixing.

\end{document}